\documentclass[12pt]{article}
\usepackage{counter-examples}
\begin{document}
\title{Counterexamples to the Theory\\
of Thermodynamics With Feedback\\
Proposed By Sagawa and Ueda}

\author{Robert R. Tucci\\
        P.O. Box 226\\
        Bedford,  MA   01730\\
        tucci@ar-tiste.com}

\date{\today}
\maketitle
\vskip2cm
\section*{Abstract}
We give some counterexamples
to the theory of thermodynamics
with feedback that has been proposed
by Sagawa and Ueda. Our
counterexamples are evidence that
 their theory, in its present form,
 is seriously flawed.
\newpage
\section*{}

In a string of papers
(Refs.\cite{Sag-U-07} to \cite{Sag-U-12b})
dating from 1997 to 2012,
Sagawa, Ueda and coworkers (S-U)
have proposed a theory
for thermodynamics with feedback.
Their work has
generated one published
comment in Ref.\cite{crit1}
to which they have replied
in Ref.\cite{response-crit1}.
This note
objects to their theory
on different grounds than
Ref.\cite{crit1}.

Numerous times throughout
Refs.\cite{Sag-U-07} to \cite{Sag-U-12b},
S-U define what they call the
``classical-quantum information" as

\beq
I_{QC}=
I(\rho_1:X)= S(\rho_1)-\sum_k p_k S(\rho_2^{(k)})
\;,
\eeq
and then they claim that

\beq
0\leq
 I_{QC}\leq H(\{p_k\})
\;.
\label{eq-wrong-ineq}
\eeq
For
instance, they claim to prove Eq.(\ref{eq-wrong-ineq})
in Ref.\cite{Sag-U-07},
Eqs.(17) and (18).
Unfortunately,
their proof of Eq.(\ref{eq-wrong-ineq})
is wrong;
both of these bounds can
be violated.
Below we will give an
example for which $I_{QC}<0$
and an example for which $I_{QC}>H(\{p_k\})$.

Henceforth,
we will denote classical random variables
by underlined letters.
In our two examples,
all density matrices are
simultaneously diagonal
and
thus correspond to classical
probability distributions.
Let
$\rho_2$
correspond to a probability
distribution  $P_{\rvx_2}$,
$\rho_1$ to $P_{\rvx_1}$,
and
$\rho_2^{(k)}$ to $P_{\rvx_2|\rvk}(\cdot|k)$
for all $k$.
In this case,\footnote{Note
that $I_{QC}\neq H(\rvx_1) - H(\rvx_1|\rvk)
=H(\rvx_1:\rvk)$.
The mutual information $H(\rvx_1:\rvk)$
does satisfy $0\leq H(\rvx_1:\rvk)\leq H(\rvk)$.}

\beq
I_{QC}= H(\rvx_1) - H(\rvx_2|\rvk)
\;.
\eeq

In both of our examples,
we will assume that the probability
distribution $P_{\rvx_2,\rvk,\rvx_1}$
can be described
by the Markov chain
$\rvx_2\larrow \rvk \larrow \rvx_1$
where $x_2,k,x_1\in \{0,1\}$.
Let
$h(\alpha)=
-\alpha \ln \alpha
-(1-\alpha) \ln(1-\alpha)$
for $\alpha\in [0,1]$.

\begin{itemize}
\item {\it Example for which $I_{QC}<0$.}

Suppose

\beq
P_{\rvx_2|\rvk}=\left[\begin{array}{cc}
\frac{1}{2}&\frac{1}{2}\\
\frac{1}{2}&\frac{1}{2}
\end{array}\right]
\;,\;\;
P_{\rvx_1}=\left[\begin{array}{c}1\\0\end{array}\right]
\;.
\eeq
Then $H(\rvx_1)=0$ and
$H(\rvx_2|\rvk)=h(\frac{1}{2})=\ln 2$
so $I_{QC}= -\ln 2$.

\item {\it Example for which $I_{QC}> H(\rvk)$.}

Suppose

\beq
P_{\rvx_2|\rvk}=\left[\begin{array}{cc}
1&0\\
0&1
\end{array}\right]
\;,\;\;
P_{\rvk|\rvx_1}=\left[\begin{array}{cc}
0&0\\
1&1
\end{array}\right]
\;,\;\;
P_{\rvx_1}=\left[\begin{array}{c}
\frac{1}{2}\\
\frac{1}{2}\end{array}\right]
\;.
\eeq
Then $H(\rvx_1)=h(\frac{1}{2})=\ln 2$ and $H(\rvx_2|\rvk)=0$
so $I_{QC}=\ln 2$. But $P_\rvk=
P_{\rvk|\rvx_1}P_{\rvx_1}=
\left[\begin{array}{c}
0\\
1\end{array}\right]$
so $H(\rvk)=0$.
\end{itemize}

In Ref.\cite{Sag-U-11}, Eqs.(30) and (31),
S-U define
quantities $I$ and $I_c$.
Then they prove
the following
 generalization of the Jarzynski equality:

\beq
\av{e^{-\sigma - I_c}}=1
\;.
\label{eq-wrong-jar}
\eeq
(see Ref.\cite{Sag-U-11}, Eq.(66)).
 They conjecture (but do not prove)
 that $\av{I}\approx \av{I_c}\approx I_{QC}$
in some sense.
But is this really a good approximation?
Note that $\av{I}$ and $\av{I_c}$
are always non-negative whereas
we've proven that $I_{QC}$
can be negative.
Note also that
Eq.(\ref{eq-wrong-jar})
implies that $\av{\sigma + I_c}\geq 0$.
If we replace $\av{I_c}$ by $I_{QC}$
in this inequality, we get
$\av{\sigma} + I_{QC}\geq 0$
which
is violated if the
entropy production $\av{\sigma}$
equals zero and $I_{QC}<0$.
S-U may have proven
$\av{\sigma + I_c}\geq 0$,
but they
never give a proof or even
a good reason
why we should expect that
$\av{\sigma} + I_{QC}\geq 0$.
And a violation
of $\av{\sigma} + I_{QC}\geq 0$
means there can be no
generalization of the Jarzynski equality
similar to Eq.(\ref{eq-wrong-jar})
but for
$I_{QC}$ instead of
 $I_c$.

Ref.\cite{Sag-U-12b}
generalizes the theory of
 Ref.\cite{Sag-U-07}
from a uni-partite to a bi-partite
system. Ref.\cite{Sag-U-12b}
 reduces to the uni-partite case
if one of the two particles is constrained to occupy
a single pure state. Thus,
all of the above objections also
apply to Ref.\cite{Sag-U-12b}.

In Ref.\cite{Tuc-max-dem},
I describe my own theory
for generalizing thermodynamics
so that it applies to processes with
feedback.
I have profitted immensely from reading
the wonderful work of S-U.
My theory agrees very much in spirit with
the S-U theory,
but differs from it in some important
details. Of course,
whether
the objections to the S-U theory
raised in this note
are valid or not
is totally independent
of the contents of Ref.\cite{Tuc-max-dem}.

\end{document}